\documentclass[prb,preprint]{revtex4-1} 

\usepackage{amsmath}  
\usepackage{amsfonts} 
\usepackage{graphicx} 
\usepackage{braket}

\begin{document}

\title{A Versatile Hong-Ou-Mandel Interference Experiment in Optical Fiber for the Undergraduate Laboratory}

\author{Cyrus Bjurlin and Theresa Chmiel\\University of Minnesota, School of Physics and Astronomy, Minneapolis, Minnesota}

\date{\today}

\begin{abstract}
Hong-Ou-Mandel (HOM) interference is a quantum optics laboratory experiment that has recently become more accessible to undergraduate students. The experiment consists of two identical photons simultaneously entering a non-polarizing beam splitter. The wavefunctions destructively interfere and the photon pairs bunch (both exit the same output) at the outputs whereas classically they are equally likely to exit different outputs. Due to the precision needed to achieve indistinguishability, setup and alignment of this experiment is often considered to be too difficult and time consuming to be appropriate for an undergraduate lab, with an end goal of merely demonstrating the HOM interference dip. Here, we present an alternative optical fiber-based apparatus that gives a consistently reproducible experiment with interference occurring in a fused-fiber coupler instead of a traditional beam splitter. We use a commercially available fiber coupled biphoton source that requires minimal alignment and increases coherence length of the interference. In addition, our biphoton source provides direct temperature based control of the frequency degeneracy of the photon pairs produced, allowing for students to investigate physical properties of HOM interference such as coherence length and interference visibility. Through use of standard opto-mechanical parts combined with the commercially available fiber integrated biphoton source and laser, our apparatus is a middle ground between built-from-scratch and pre-aligned setups.
\end{abstract}

\maketitle

\section{Introduction} 
The nature of interference at the quantum scale is a source of confusion for many physics students. One such phenomenon is known as Hong-Ou-Mandel (HOM) interference and was first demonstrated by Hong et al. in 1987.\cite{hong1987} HOM interference consists of two individual photons entering the two input ports of a 50:50 non-polarizing beam splitter. When two photons have identical frequency, polarization, arrival time, and position in space, we say they are quantum mechanically indistinguishable. When photons entering the beam splitter are indistinguishable, HOM interference occurs. Classically, you would expect to see the two photons exiting at each of the two output ports 50 percent of the time. When interference happens, the photons are more likely to exit the same output port and few to no coincidences are measured. When this happens, we say that photons "bunch" at the output. This makes HOM interference a very intuitive demonstration of quantum optics.

HOM interference is realizable in an undergraduate laboratory,\cite{dibrita2023,ourjoumtsev2015,carvioto2012} however it is usually implemented as a demonstration. The reason for this is that alignment of an HOM apparatus can be difficult and time consuming for an undergraduate student to perform. In order to achieve indistinguishable photons, the two input beams of the beam splitter must be aligned horizontally and vertically so that they overlap when exiting the beam splitter and the two photons must enter the beam splitter almost simultaneously (so that their spatial modes overlap). Furthermore, the common method of producing photon pairs using spontaneous parametric down conversion in $\beta$-Barium Borate crystals produces photon pairs that have an HOM coherence length on the order of 10 microns,\cite{hong1987,ourjoumtsev2015} which is a level of precision that often requires tedious alignment steps at the beam splitter before aligning the actual interference beams. Due to these many limitations, investigating anything more than initial interference would be too intensive a project for most undergraduate teaching labs to find worthwhile.

We address these challenges by presenting an apparatus that not only allows for easier alignment, but also provides opportunities to investigate other physical properties of HOM interference. Through the use of both commercially available pre-aligned parts as well as parts that must be aligned by the student, we find a middle ground between buying a full setup off the shelf and having to meticulously align every part of the apparatus. First, the majority of the apparatus consists of fiber optic cables which minimizes the alignment procedures necessary. Second, we employ a fiber integrated photon pair source that produces pairs with coherence length in the hundreds of microns, making effective simultaneity easier to achieve as well as eliminating the tedious alignment of the traditional down conversion setup. Third, the photon pair source can be tuned to adjust the difference in frequency between the two photons produced as well as the rates of photon pair production. This gives more experimental parameters that can be investigated in relation to the behavior of HOM interference. The remainder of the apparatus however must be built and aligned by the student giving them valuable experience with opto-mechanical setup. The combination of commercially available and do-it-yourself parts makes this apparatus ideal for a longer advanced undergraduate experiment as it would be difficult to complete in a week, but it isn't so difficult that a student couldn't achieve it given time.

\section{Apparatus}
The interfering photon pairs are produced using a phenomenon known as spontaneous parametric down conversion (SPDC). SPDC occurs when a high-power pump laser hits a crystal with nonlinear optical properties. Pump photons are converted into two photons known as the signal and idler, with the requirement that $\omega_i + \omega_s = \omega_p$ where $\omega_{i,s,p}$ are the frequencies of the idler, signal, and pump respectively. Because the signal and idler are produced at the same time, when we measure photon counts over a sufficiently short time we can assume that if we measure one count at each of two outputs, one is the signal and one is the idler so we have a coincidence.  

We produce photon pairs using collinear type II SPDC, meaning the signal and idler are polarized orthogonally and emitted in the same direction of travel as the pump photon. The polarization of the emitted photons is what distinguishes type II SPDC from type I or 0 where the signal and idler have the same polarization. Specifically, the nonlinear crystal used is periodically poled potassium titanyl phosphate (PPKTP).\cite{fedrizzi2007} 

\begin{figure}[h!]
\centering
\includegraphics[width=4.8in]{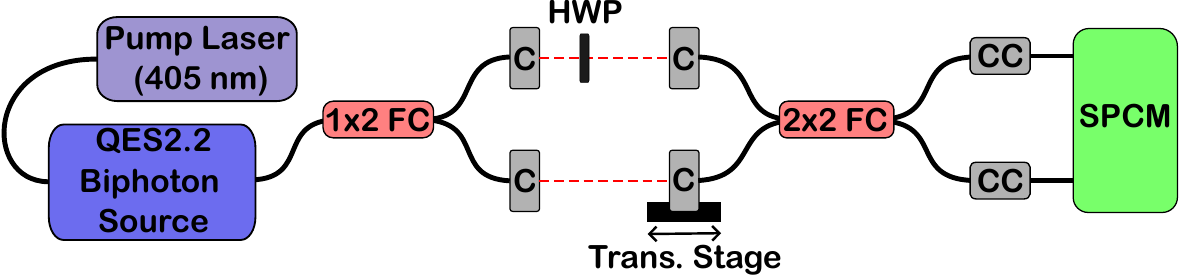}
\caption{A diagram of the interference apparatus. 1x2 FC is the one input two output fused-fiber coupler, C labels the 810 nm collimators, HWP is the 810 nm half-wave plate, 2x2 FC is the two input two output fused-fiber coupler, CC labels the coupler cages which filter visible light, and SPCM stands for single photon counting module which records both single photon and coincidence rates for the two fiber coupler outputs. The solid connecting lines in the diagram are single mode polarization maintaining fiber optic cable and the dotted line is the path of the photons in air.}
\label{apparatuscartoon}
\end{figure}

A diagram of the full apparatus is given in Fig \ref{apparatuscartoon}. The pump laser (Ondax, LMFC-405) is a 405 nm laser diode coupled to single-mode polarization maintaining (PM) fiber. All of the fiber used in the interference apparatus is single-mode and polarization maintaining in order to prevent modal and polarization dispersion\cite{thyagarajan2007}. The biphoton source (Qubitekk, QES2.2) is coupled to single-mode PM fiber and is directly connected to the pump laser via a ferrule connector/physical contact (FC/PC) fiber mating sleeve (ThorLabs, ADAFC2). This biphoton source consists of a temperature controlled PPKTP crystal with a spectral filter to remove the pump beam from the output. The photon pairs produced then have wavelengths close to 810 nm, with wavelengths of exactly 810 nm for indistinguishable output. The serial emulator Termite\cite{riemersma2019} is used to interface with the biphoton source and control the PPKTP crystal temperature. Due to temperature dependence of PPKTP's optical properties,\cite{fedrizzi2007} the temperature control in the biphoton source allows direct control of the frequency degeneracy of the output. The photon pairs are orthogonally polarized when they leave the biphoton source, where they enter a one input two output fused-fiber coupler (Thorlabs, pfc780f)---labeled 1x2 FC in Fig \ref{apparatuscartoon}---which separates them between two outputs based on their polarization. 

\begin{figure}[h!]
\centering
\includegraphics[width=4.8in]{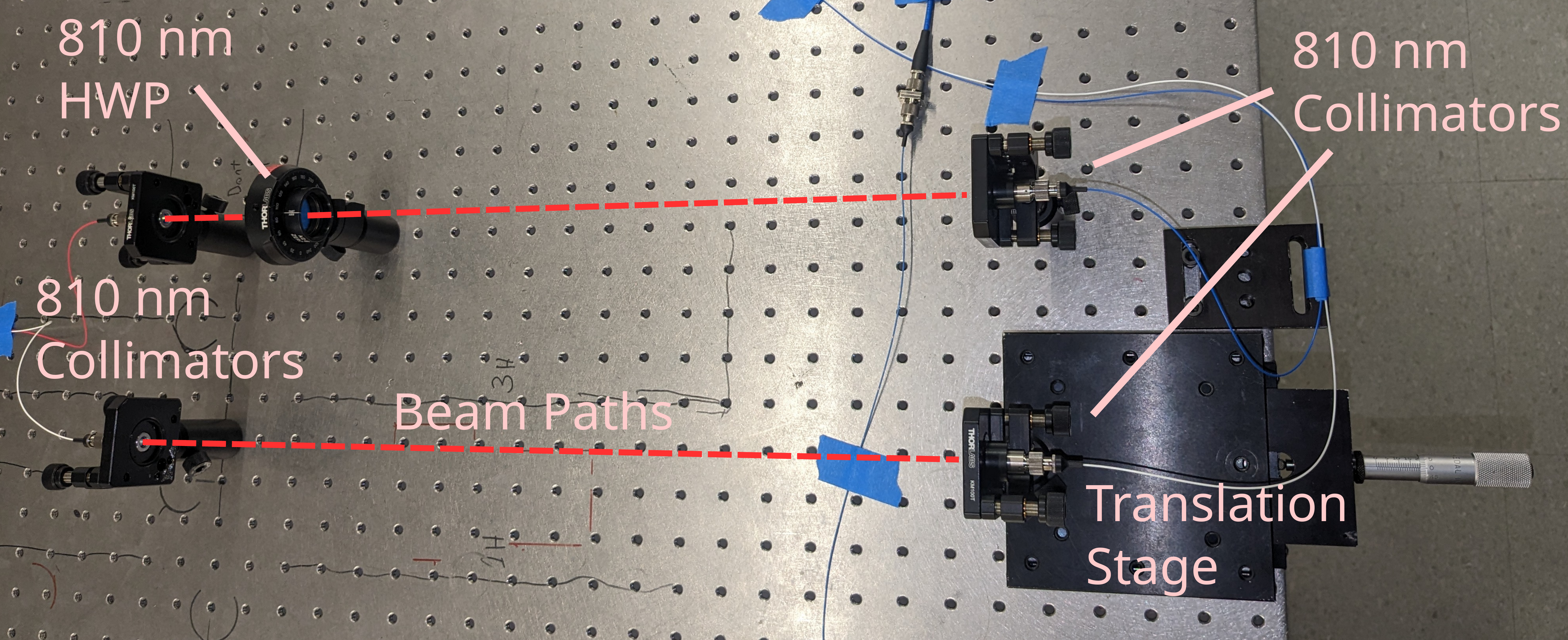}
\caption{A labeled picture of the fixed and adjustable delay lines with the beam paths as dashed lines. The 810 nm half wave plate is in the fixed delay line. The translation stage is used to change the length of the adjustable delay line.}
\label{delaypic}
\end{figure}

The two outputs of the 1x2 coupler are connected using FC/PC connectors to two collimators consisting of a fixed focus collimating lens (Thorlabs, F280FC-780) mounted (Thorlabs, KM100T) with xy tilt control. These two collimators start the in-air section of the apparatus. This section consists of two delay lines, one is fixed and one is adjustable. Details of the two delay lines are shown in Fig \ref{delaypic}. Each delay line has another identical collimator at its end. The collimators are all oriented so that the photons leave and enter with vertical polarization and an 810 nm half-wave plate is inserted into the fixed delay line to precisely match the polarization in the adjustable delay line. Note that although the photons are initially orthogonally polarized, they both leave the 1x2 fused-fiber coupler vertically polarized due to the orientation of the connector keys at the outputs. The adjustable delay line is terminated with a collimator mounted to a translation stage (Thorlabs, PT1) which adjusts the optical path-length difference of the photon pair as they enter the next section of the apparatus. The translation stage has a resolution of 0.02 mm and a range of 25 mm which is precise enough to show the shape of the HOM dip but more than large enough to capture the entire dip which is on the order of 1 mm wide.

Because the output of the biphoton source is not visible light, it is easiest to first roughly align the two delay lines using a visible laser and maximizing counts through the delay lines. We use a 633 nm HeNe laser (Melles Griot, 05-lhp-151) which is coupled into a multi-mode fiber (Thorlabs, M31L01) using a collimator with both XY tilt and translation control consisting of a mount (Thorlabs, K6XS) and fixed focus collimating lens (Thorlabs, F220FC-B). More precise alignment is then done using the 810 nm photons.

The collimators at the end of the delay line are connected to a two input two output fused fiber coupler (OF Link, FPMC-810-22-50-PM-C-05-FP-B), which is labeled 2x2 FC in Fig \ref{apparatuscartoon}. Unlike the first coupler, this one takes the two photons in both polarized vertically and it sends each with equal probability to either of the two outputs. This is the analog of a nonpolarizing beam splitter in other HOM setups. The outputs of the 2x2 coupler are connected to two coupler cages, labeled CC in Fig \ref{apparatuscartoon}, that use longpass filters (Thorlabs, FGL780) to filter out visible light before the remaining photons are sent to the single photon counting module (Excelitas, SPCM-AQRH-13-FC), abbreviated to SPCM, which creates a pulse that is sent to the field porgrammable gate array, or FPGA, (Digilent Nexys3 100 MHz system clock) which records both the counts at each output as well as  the coincidences between them.

The data collected by the single photon counting module and FPGA counting module is sent to a computer running a Labview program which shows live plots of single photon and coincidence rates. The program records the raw data to a comma separated values (CSV) file while more detailed analysis is done in Excel or Matlab.

\section{Theory}
\subsection{Fused Fiber Coupling}
HOM interference was first demonstrated in a non-polarizing beam splitter and even the undergraduate lab demonstrations that use fiber optics often use a beam splitter to achieve the actual interference.\cite{hong1987,dibrita2023,weihs1996} However in this setup, the HOM interference occurs within a fused fiber coupler. It is therefore necessary to understand what exactly a fused-fiber coupler does to a photon at the quantum mechanical level in order to motivate the theory of the HOM effect.

Fused-fiber couplers are traditionally understood as an application of the evanescent electric field phenomenon.\cite{pal2003} However, it is equivalent to consider the wave-function of a photon in a single-mode fiber. A single-mode fiber consists of a core several microns across surrounded by a cladding of lower refractive index.\cite{thyagarajan2007} This effectively forms a cylindrical potential well. Due to continuity requirements, the wave-function extends slightly out of the core of the fiber into the cladding. When the cores of two fibers are brought very close together with only a small amount of cladding between them, the system turns into two potential wells with a thin barrier, where the wave-function of a photon in one well can tunnel into the other. If the wave-function remains in this potential, the probability density will oscillate back and forth between the two wells.\cite{peacock2006}

A fused fiber coupler then consists of two fibers, usually single-mode, which are fused together in the middle to create the potential described above. By adjusting the length of the coupling region, a manufacturer can control the time that the photon remains in this potential and effectively control the probability of finding the photon at either of the outputs. The fiber coupler is symmetric, so the the probability of staying or switching fibers is the same for both inputs. From a quantum mechanical perspective, the action of a fused-fiber coupler on an input state is the same as the action of a beam-splitter on an input state.\cite{weihs1996} Therefore, just as with a beam-splitter, we can represent the action of the fused-fiber coupler with the unitary 2x2 matrix operator
\begin{equation}
    \hat M = 
    \begin{pmatrix}
        t & r \\
        -r & t \\
    \end{pmatrix}
\label{splittermatrix}
\end{equation}
where $t$ is the transmission coefficient and $r$ is the reflection coefficient such that $t^2 + r^2 = 1$. The negative sign on the bottom left $r$ is added to represent a phase change that is required due to energy conservation.\cite{makarov2022}

\subsection{HOM interference}
Although our experiment occurs in a fused-fiber coupler, it is instructive to consider exactly what HOM interference is in a beamsplitter before considering a more mathematically rigorous treatment of the coupler interference. When two photons come to the two inputs of a 50:50 non-polarizing beam splitter, they can behave in one of four ways. 
\begin{enumerate}
    \item One photon transmits and one reflects, leading to both photons exiting one output.
    \item The other photon transmits and the first reflects leading to both exiting the other output.
    \item Both photons transmit and exit at opposite outputs.
    \item Both photons reflect and exit at opposite outputs.
\end{enumerate}
Classically we would expect to see all four options equally often. However, when the photons are indistinguishable, HOM interference causes options 1 and 2 to become much more likely than options 3 and 4 which in turn leads to a decrease in coincidences measured between the two outputs.

\begin{figure}[h!]
\centering
\includegraphics[width=4.8in]{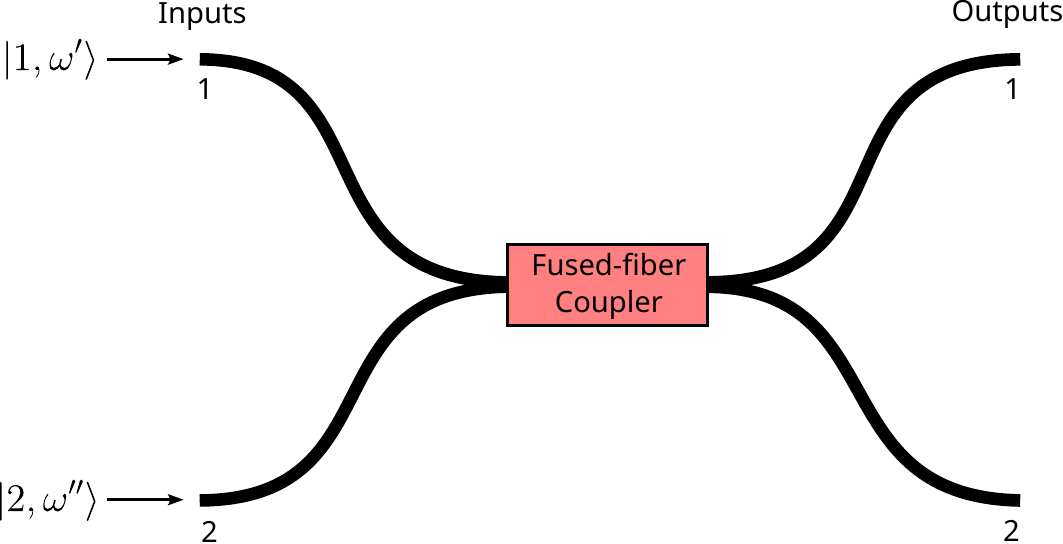}
\caption{The two-photon state enters the fused-fiber coupler. Here a photon of frequency $\omega'$ enters at input 1 and a photon of frequency $\omega''$ enters at input 2. The fused-fiber coupler then sends photon 1 to outputs 1 or 2 with probabilities $t^2$ or $r^2$ respectively and sends photon 2 to outputs 1 or 2 with probabilities $r^2$ or $t^2$ respectively.}
\label{couplerdiagram}
\end{figure}

In our experiment we use a fused-fiber coupler which, from Eq. \ref{splittermatrix}, behaves similarly to a beam splitter. With the action of the coupler defined, the purely quantum mechanical phenomenon of HOM interference can be described. Similar to the formulation of Weihs et al.,\cite{weihs1996} the state of the two-photon system entering the coupler (Fig \ref{couplerdiagram}) is described by the double integral
\begin{equation}
\label{twophotonstate}
    \ket{\Psi_{in}}=\int\int d\omega'd\omega''\zeta(\omega',\omega'')\ket{1,\omega'}\ket{2,\omega''},
\end{equation}
where
\begin{equation}
    \zeta(\omega',\omega'') \propto \delta(\omega'+\omega''-\omega_p)f_{\omega'_c}(\omega')f_{\omega''_c}(\omega'').
\end{equation}
Here, $\ket{1,\omega'}$ represents a photon of frequency $\omega'$ entering the coupler at input 1, $f_{\omega'_c}(\omega')$ is the single photon frequency distribution, and $\delta$ is the Dirac delta distribution, added to adhere to the constraint that $\omega'+\omega''$ is equal to the pump frequency $\omega_p$. 
Although SPDC in PPKTP crystal produces a $sinc^2$ spectrum,\cite{fiorentino2007}  a gaussian distribution can be used as an accurate approximation of the $sinc^2$ function.\cite{dattoli1998,baghdasaryan2022,fedorov2009} Therefore $f_{\omega'_c}(\omega')$ is assumed to be a sharply peaked gaussian distribution with center frequency $\omega'_c$. 

The interference is then given by the action of  the matrix operator $\hat M$ on the photons at the two inputs:
\begin{equation}
\label{coupleraction}
    \ket{\Psi_{out}} = (\hat M\otimes\hat M)\ket{\Psi_{in}}.
\end{equation}
There is a more in depth explanation of the action of such a matrix on the two-photon state in Appendix A as well as calculation of probabilities of outcomes. Using the method of Appendix A and then integrating over all wavelengths we can find probabilities of the different outcomes for the final state $\ket{\Psi_{out}}$. The details of these calculations can be found in Campos et al.\cite{campos1990} and produce the probabilities
\begin{eqnarray}
    P_{out}(2,0) &=& P_{out}(0,2) = t^2r^2[1+\Phi] \\
    P_{out}(1,1) &=& t^4 + r^4 -2t^2r^2\Phi
\label{probeq}
\end{eqnarray}
where
\begin{equation}
    \Phi = \int\int\zeta(\omega',\omega'')\zeta^*(\omega'',\omega')d\omega'd\omega''.
\end{equation}
When the single photon distributions are gaussian as in our case, $\Phi$ is evaluated as\cite{campos1990}
\begin{equation}
\label{overlap}
    \Phi = e^{\frac{-\omega_d^2t_c^2}{8}}\times e^{\frac{-2\Delta x^2}{c^2t_c^2}},
\end{equation}
where $c$ is the speed of light, $\omega_d = \lvert\omega'_c - \omega''_c\rvert$, $t_c$ is the reciprocal of the spectral width of the two photon distribution, and $\Delta x$ is the optical path length difference between the photons arriving at the coupler. This equation assumes the photons to be identical in polarization. When both $\omega_d = 0$ and $\Delta x = 0$, the photons are quantum mechanically indistinguishable and $P_{out}(1,1) = (t^2-r^2)^2$ achieves full destructive interference when $t^2 = r^2 = 1/2$. Plotting $P_{out}(1,1)$ vs $\Delta x$ gives the characteristic ``HOM dip'', an example of which is shown in Fig \ref{bigdipfig}. 

Note that Eq. \ref{overlap} has a factor dependent on $\omega_d$ as well as a factor dependent on $\Delta x$. Therefore we can also plot an ``HOM dip" with the relationship between coincidence counts and $\omega_d$. We produce our photon pairs through collinear type II spontaneous parametric down conversion (SPDC).\cite{qubitekk2017} This specific type of down conversion is highly temperature dependent with slight changes in the temperature of the down converting crystal affecting $\omega_d$ for the photon pairs produced.\cite{steinlechner2014} A unique aspect of this experimental setup is the ability to control the crystal temperature in the bi-photon source.

We define the central temperature $T_c$ to be the crystal temperature at which the photon pairs have the same wavelength or $\omega_d = 0$. Fluctuations around $T_c$ will induce fluctuations in the width of the photon distribution $\omega_d$. Defining $\Delta T = \lvert T_c - T\rvert$, we can say that $\Delta T \propto \omega_d$. \cite{steinlechner2014} Therefore it is equivalent to plot a dip with temperature as the independent variable. We can do this by measuring and plotting interference visibility at varying temperatures.

\subsection{Treatment of Accidentals}
The data collection in this experiment relies heavily on counting coincidences at two detectors. The inherent flaw in doing this is that we can't truly tell when two events are exactly simultaneous. We define a coincidence between two detectors to be when both detectors measure a count within the same time span, called the coincidence window and denoted $\tau_c$. When the gap between individual counts at a detector is much larger than $\tau_c$, this definition of coincidence is very effective at measuring actual instances of simultaneity. However, there will still be some accidental coincidences that don't come from photon pairs. The rate of accidental coincidences $R_{acc}$ is given by\cite{pearson2010}
\begin{equation}
    R_{acc} = \tau_cR_AR_B
\end{equation}
where $R_A$ and $R_B$ are the rates measured at detectors A and B respectively. Using this relationship we calculate accidental coincidence rates and subtract them from the overall data.
\section{Sample Data and Analysis}
\begin{figure}[h!]
\centering
\includegraphics[width=5.5in]{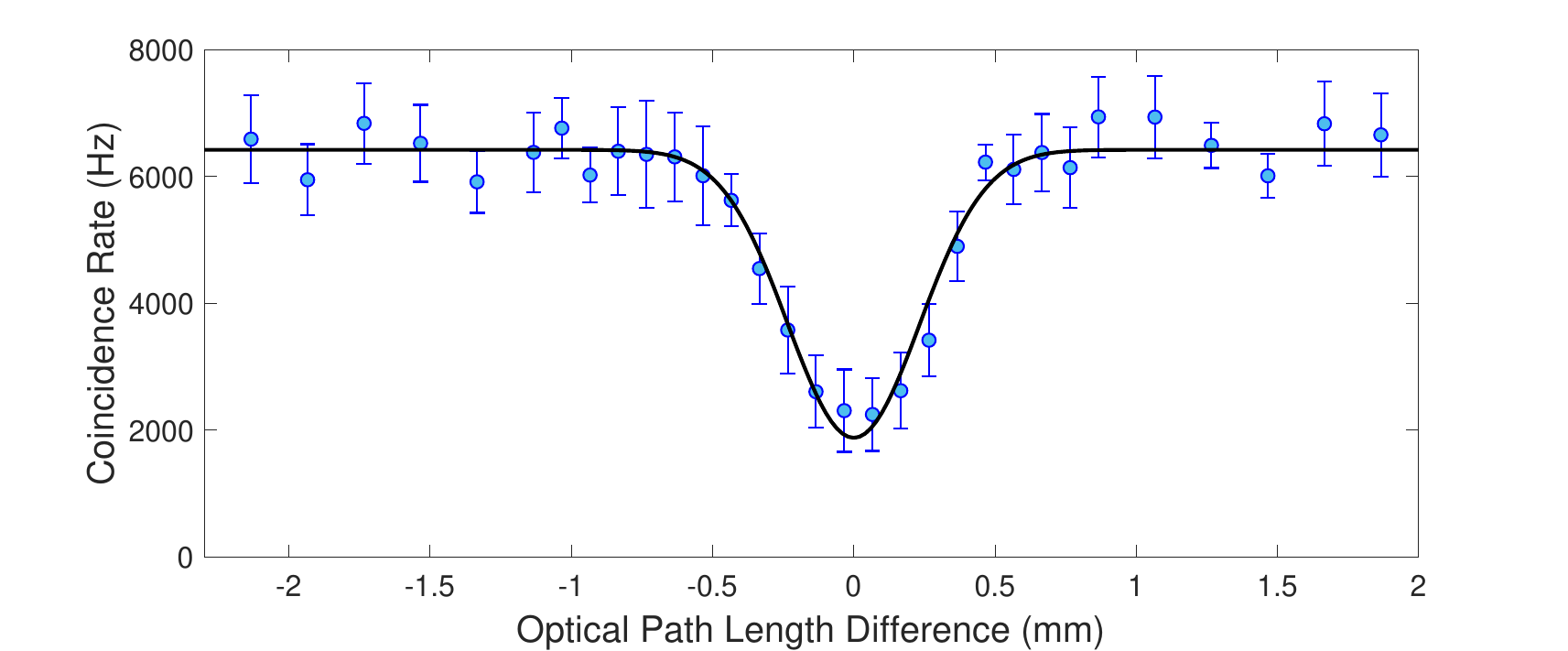}
\caption{Coincidence rate plotted against optical path length difference (OPD). Crystal temperature for the measurements is $(35.93\pm0.05)^{\circ}\textrm{C}$. Error bars correspond to standard error of the averaged data. Accidental coincidence rates are subtracted and the data is fitted to an inverted gaussian distribution with equation $y = (6420\pm160) \textrm{Hz} - (4541\pm465)\textrm{Hz} \times e^{(-9.21\pm1.98)\textrm{mm}^{-2} x^2}$ This shows the characteristic ``HOM dip'', where the strong interference occurs when the OPD is equal to 0.}
\label{bigdipfig}
\end{figure}

\begin{figure}[h!]
\centering
\includegraphics[width=4.8in]{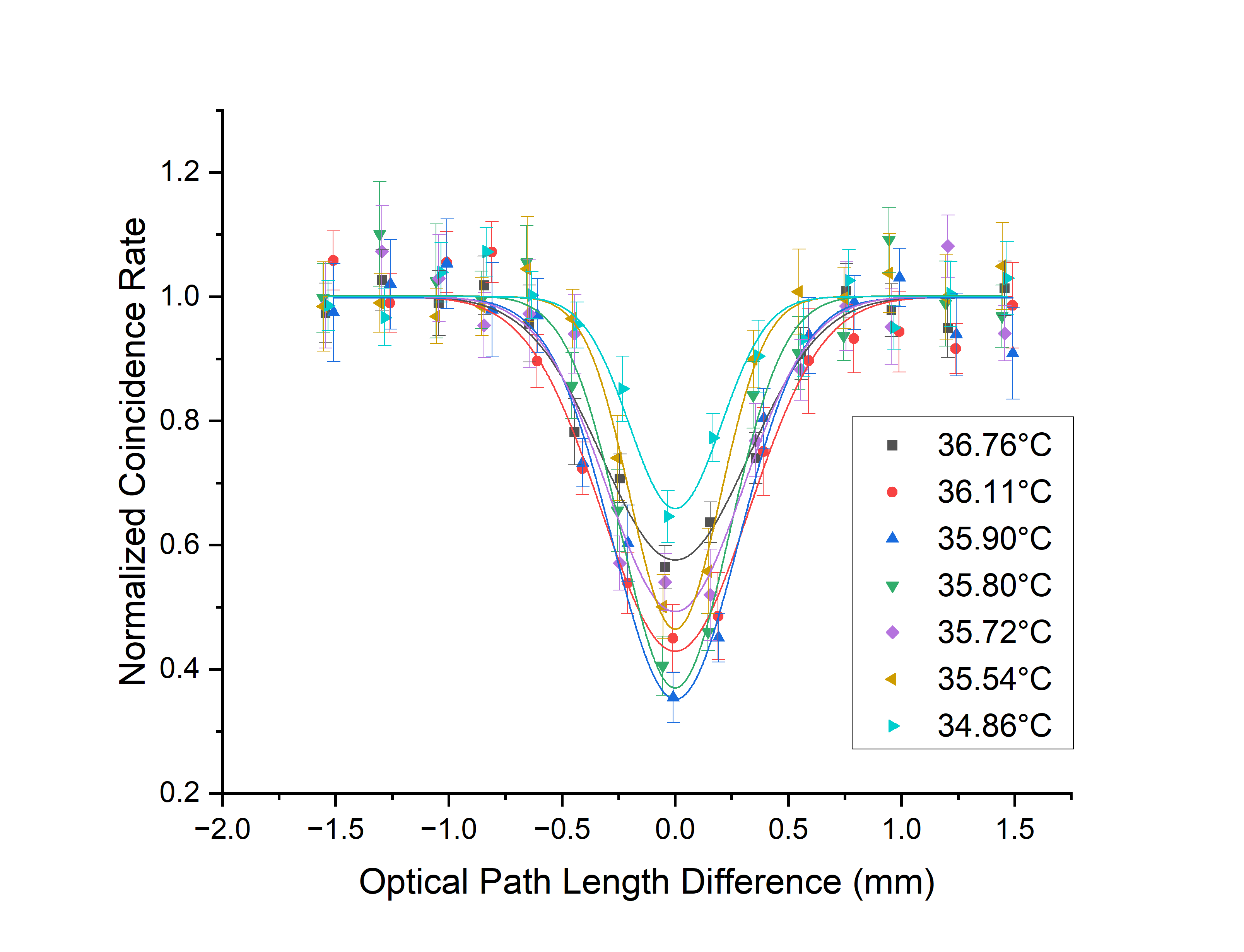}
\caption{Coincidence counts plotted against optical path length difference (OPD) for seven different crystal temperatures. The data sets are each fitted to an inverted gaussian distribution (Fig. \ref{bigdipfig}) and are normalized with respect to the average coincidence counts outside of the dip with accidentals removed.}
\label{hom5dips}
\end{figure}

In collecting the sample data, coincidence counts and single counts were recorded using a Labview program and analyzed using Matlab. The Labview program records the coincidence counts every 0.3 seconds using a coincidence gate time of 10 ns and outputs the coincidence rate over that 0.3 second period. For each data point in Fig. \ref{bigdipfig} 60 seconds of coincidence rate data or roughly 200 individual coincidence rates were recorded. The value of each data point is then the average of those coincidence rates. For the dips in Fig. \ref{hom5dips}, the data points were each averages over 30 seconds or 100 coincidence rates. The error bars represent the standard deviation of the mean.

Merely demonstrating the existence of the dip is often the end goal of an undergraduate HOM experiment. An example of that result is shown in Fig. \ref{bigdipfig}. Coincidence rates are measured in 0.1 mm increments near the point of zero optical path length difference and 0.2 mm increments outside the center range. After the accidental coincidence rates are subtracted out, the data is fitted to an inverted Gaussian distribution.

The depth of the dip, also known as the interference visibility, does not reach zero as would be expected for perfect indistinguishability. The main reason for this is not that the photons aren't indistinguishable but that the two input two output fused-fiber coupler where the interference occurs does not have equal probability of sending a photon to either output. It can be shown using Eq. \ref{probeq} that the visibility of the dip is given by
\begin{equation}
\label{visequation}
    V = \frac{2r^2t^2}{r^4 + t^4}
\end{equation}
where $t^2$ and $r^2$ are the transmission and reflection probabilities respectively. $V = 0$ when there is no dip and $V = 1$ when the dip goes fully to zero coincidences. We measured the transmission and reflection of the fused-fiber coupler for indistinguishable photons and found that $t^2 = 0.466\pm0.003$ and $r^2 = 0.534\pm0.003$. Using Eq. \ref{visequation}, the maximum visibility that could be recorded using this apparatus is $0.991\pm0.004$. The data of Fig. \ref{bigdipfig} does not quite reach this mark, but we achieve interference visibility of $0.707\pm0.072$, which passes the $50\%$ mark necessary to distinguish classical and quantum HOM interference.\cite{na2020}

We demonstrate interference and confirm that the data follows an inverted gaussian model in Fig. \ref{bigdipfig}. This is interesting on its own, but we can investigate more characteristics of HOM interference.

By repeating the measurement process of Fig. \ref{bigdipfig} for multiple scenarios, we can investigate how frequency degeneracy effects the HOM dip. Because of the temperature control in the QES2.2 biphoton source, we are able to take measurements at non-degenerate wavelengths. In Fig. \ref{hom5dips}, coincidence rates vs optical path length difference is plotted for seven different temperatures near the degenerate temperature $T_c$. For these graphs, data points were taken every 0.20 mm. 

The dip is deepest at the biphoton source temperature $35.90 ^{\circ}\textrm{C}$ (Fig. \ref{hom5dips}). By finding the temperature at which the visibility will be the greatest, we can then use that temperature to plot our best dip. This data shows the uniqueness of this apparatus; measurements for different degeneracies would not be possible for most implementations of HOM interference in an undergraduate lab.\cite{dibrita2023,ourjoumtsev2015}

\begin{figure}[h!]
\centering
\includegraphics[width=4.8in]{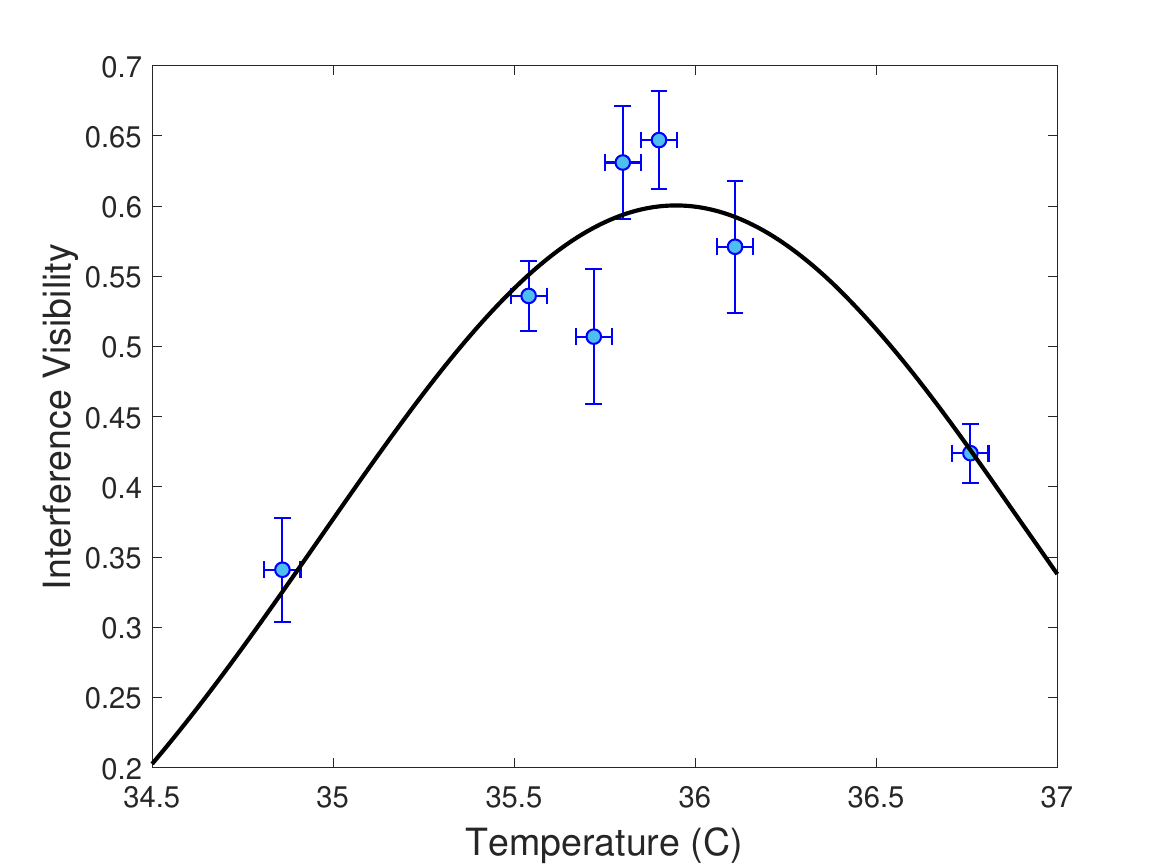}
\caption{Dip visibility plotted against PPKTP crystal temperature. This data is assumed to follow Gaussian distribution as in Eq. \ref{overlap} and is fitted to a model of that form. Horizontal error bars represent estimated crystal temperature fluctuations, vertical error bars represent a 67\% confidence interval in dip visibility. The fit has a reduced $\chi^2$ of 1.5 and and adjusted $R^2$ of 0.86.}
\label{fwhmfitfig}
\end{figure}

For a more precise investigation of the relationship between degeneracy and visibility, visibility of the dips is plotted against PPKTP crystal temperature in Fig. \ref{fwhmfitfig}. This will allow us to precisely find $T_c$. 

This data in Fig. \ref{fwhmfitfig} is fitted to a Gaussian model. The data in Fig. \ref{fwhmfitfig} exemplifies the utility and uniqueness of this apparatus as temperature control is not often found in undergraduate HOM labs. From the fitting data the central temperature was calculated to be $(35.95 \pm 0.13)\,^{\circ}\textrm{C}$.

\section{Conclusion}
We presented a versatile and user-friendly apparatus for investigating HOM interference in an undergraduate laboratory. The optical fiber-based apparatus allows for easier investigation of coherence length and interference visibility for both distinguishable and indistinguishable photon pairs. As an example of the possible experiments that could be done, we showed a relationship between interference visibility and PPKTP crystal temperature. 

We demonstrate interference and show that the HOM dip follows an inverted gaussian profile (Fig. \ref{bigdipfig}). We also find the degenerate temperature $T_c$ and show the effect of crystal temperature on interference visibility.

This apparatus is well suited for implementation in an undergraduate lab. The multitude of fiber optic components make alignment and setup more intuitive and less tedious for a student who may be less comfortable with experimental optics, while the alignment of the in-air delay lines still gives experience using more traditional optomechanical equipment, making this lab a perfect middle ground between off the shelf setups and setups built from the ground up. Furthermore, the use of fiber optics, specifically the fused fiber couplers, exposes a student to more specialized physical phenomena, as the physics of fiber optics and SPDC are not commonly included in undergraduate curricula.\cite{national2013} 

These types of quantum experiments for undergraduates are being implemented by other labs and they can provide valuable intuition for students towards the behavior of quantum mechanics.\cite{galvez2014} Although quantum interference phenomena are common examples of the ``weirdness'' of quantum mechanics,\cite{mullin2017} undergraduates rarely get to see the specific physical characteristics of these phenomena. This apparatus provides that opportunity for HOM interference. 

Here, we investigate the relationship between temperature and visibility, however there are other possible experiments that could be done. We have a clear method of measuring the coherence length of the interference by measuring the FWHM. The relationship and possible dependency of coherence length on the various parameters of interference could be investigated.

Another experiment that could be done involves the polarization of the interfering photons. In this paper we have exclusively scanned over optical path length difference to create the HOM dips. However, it is equally valid to consider differences in polarization between the photon pairs instead of path length.\cite{campos1990} The behavior of the HOM dips with respect to polarization would be interesting to investigate using this apparatus and would require careful consideration of the how the changes in polarization affect the behavior of the polarization maintaining fused-fiber coupler.

\section{Author Declarations}
\subsection{Conflict of Interest}
The authors have no conflict of interest to disclose.

\section*{Appendix A: Action of the Beamsplitter Matrix on the Two-Photon State}
While this math is not vitally important to understanding this paper, it is helpful for students less familiar with these types of probability distributions to get a clearer view of what's going on. When our photons pass through the fused fiber coupler we are in fact acting on the state of Eq. \ref{twophotonstate} with the matrix operator $\hat M\otimes\hat M$ where $\hat M$ is the matric operator of Eq. \ref{splittermatrix}. Although our two-photon state is a more complicated integral over different wavelengths, we can demonstrate the action of $\hat M\otimes\hat M$ using a much simpler state.

Suppose instead of the state of Eq. \ref{twophotonstate}, our two-photon system only has one wavelength. Then it would be represented by
\begin{equation}
    \ket{\Psi_{in}}=\ket{1}\ket{2} = \ket{1}\otimes\ket{2},
\end{equation}
Where $\ket{1}$ is the photon entering input 1 and $\ket{2}$ is the other photon entering input 2. The action of the coupler operator is much simpler and gives us:
\begin{equation}
\label{matrixaction}
\begin{split}
    (\hat M\otimes\hat M)\ket{1}\ket{2} &= (\hat M \ket{1})\otimes(\hat M \ket{2})\\
    &= 
    \begin{pmatrix}
        t & r \\
        -r & t \\
    \end{pmatrix}
    \begin{pmatrix}
        1 \\
        0 \\
    \end{pmatrix}
    \otimes
    \begin{pmatrix}
        t & r \\
        -r & t \\
    \end{pmatrix}
    \begin{pmatrix}
        0 \\
        1 \\
    \end{pmatrix}\\
    &= 
    \begin{pmatrix}
        t \\
        -r \\
    \end{pmatrix}
    \otimes
    \begin{pmatrix}
        r \\
        t \\
    \end{pmatrix}\\
    & = (t\ket{1}-r\ket{2})\otimes(r\ket{1}+t\ket{2})\\
    & = tr\ket{1}\ket{1} + t^2\ket{1}\ket{2} - r^2\ket{2}\ket{1} - tr\ket{2}\ket{2}.
    \end{split}
\end{equation}
When we assume that the two photons are indistinguishable, we get 
\begin{equation}
    \ket{1}\ket{2} =\ket{2}\ket{1}
\end{equation} 
and Eq. \ref{matrixaction} becomes
\begin{equation}
   (\hat M\otimes\hat M)\ket{1}\ket{2} = tr\ket{1}\ket{1} + (t^2-r^2)\ket{1}\ket{2} - tr\ket{2}\ket{2}.
\end{equation}
Note that when $t = r = \frac{\sqrt{2}}{2}$, $t^2-r^2 = 0$ the only possible outcomes are $\ket{1}\ket{1}$ and $\ket{2}\ket{2}$ giving perfect HOM interference and no coincidences. 

However when $r \neq t$, calculating probabilities for each outcome is still quite simple. Since $r$ and $t$ are real numbers, we merely square the coefficient of each state in the wavefunction to find its probability. 


Therefore the probability of a coincidence would be $(t^2-r^2)^2$ and the probability of not measuring a coincidence would be $2t^2r^2$. These calculations are in fact exactly what is happening behind the scenes when we evaluate Eq. \ref{coupleraction} to find the probabilities of different outcomes. The only difference is this type of calculation is done at every possible wavelength pairing for the two photons and then integrated over all wavelengths. A more detailed treatment of this mechanism as well as other mathematical properties of a beam splitter can be found in Makarov, 2022.\cite{makarov2022}

\newpage
\section*{Appendix B: Parts and Costs}
\centering
\subsection*{Alignment}

\begin{tabular}{|c|c|c|c|}
    \hline
    Qty. & Part & Product Identification & Price (ea.) \\
     \hline\hline
   1 & Melles Griot HeNe Laser Head & 05 LHP 151 & \$300 - \$600 \\
     \hline
     1 & Thorlabs 6 axis collimator mount & K6XS & \$300 \\
     \hline
     1 & Thorlabs 633 nm collimating lens & F220FC-B & \$180 \\
     \hline
     1 & Thorlabs steel post & TR series & \$5 \\
     \hline
     1 & Thorlabs post holder & PH series & \$10 \\
     \hline
     2 & Thorlabs multimode patch cable & M31L01 & \$60 \\
     \hline
     2 & Thorlabs single mode patch cable & P1-780A-FC-1 & \$100 \\
     \hline
\end{tabular}

\subsection*{Photon Pair Production}

\begin{tabular}{|c|c|c|c|}
    \hline
    Qty. & Part & Product Identification & Price (ea.) \\
     \hline\hline
   1 & Ondax 480 nm diode laser & LM series & \$10,850 \\
     \hline
     1 & Qubitekk biphoton source & QES 2.2 & \$8,685 \\
     \hline
     1 & Thorlabs 1x2 PM fused fiber coupler & PFC780F & \$590 \\
     \hline
\end{tabular}

\subsection*{In-Air Delay Lines}

\begin{tabular}{|c|c|c|c|}
    \hline
    Qty. & Part & Product Identification & Price (ea.) \\
     \hline\hline
   4 & Thorlabs xy adjusting collimator mount & KM100T & \$80 \\
     \hline
     4 & Thorlabs collimating lens & F280FC-780 & \$180 \\
     \hline
     5 & Thorlabs steel post & TR series & \$5 \\
     \hline
     5 & Thorlabs post holder & PH series & \$10 \\
     \hline
     1 & Thorlabs translation stage & PT1 & \$330 \\
     \hline
     1 & Thorlabs 808 nm half-wave plate & WPH10M-808 & \$620 \\
     \hline
     1 & Thorlabs flip mount adapter & FM90 & \$95 \\
     \hline
     1 & Thorlabs manual rotation mount & RSP1 & \$100 \\
     \hline
\end{tabular}

\subsection*{Interference}

\begin{tabular}{|c|c|c|c|}
    \hline
    Qty. & Part & Product Identification & Price (ea.) \\
     \hline\hline
   1 &OF Link 2x2 pm 810 nm fused fiber coupler & FPMC-810-22-50-PM-C-05-FP-B& \$450 \\
     \hline
\end{tabular}

\subsection*{Filter Cages}

\begin{tabular}{|c|c|c|c|}
    \hline
    Qty. & Part & Product Identification & Price (ea.) \\
     \hline\hline
   2 & Thorlabs cage plate & CP33 & \$20 \\
     \hline
     2 & Thorlabs cage-compatible kinematic mount & KC1T & \$110 \\
     \hline
     4 & Thorlabs 633 nm collimating lens & F220FC-B & \$180 \\
     \hline
     2 & Thorlabs longpass filter & FGL 780 & \$50 \\
     \hline
     8 & Thorlabs cage post & SR4 & \$10 \\
     \hline
     2 & Thorlabs lens tube & SM1L10 & \$15 \\
     \hline
     2 & Thorlabs steel post & TR series & \$5 \\
     \hline
     2 & Thorlabs post holder & PH series & \$10 \\
     \hline
\end{tabular}

\subsection*{Measurement}

\begin{tabular}{|c|c|c|c|}
    \hline
    Qty. & Part & Product Identification & Price (ea.) \\
     \hline\hline
   2 & Excelitas photon counting module* & SPCM-AQRH-13-FC & \$1955 \\
     \hline
     1 & Digilent FPGA & nexys 3 & \$200 - \$400 \\
     \hline
\end{tabular}

\small{*Excelitas photon counters purchased through the ALPhA single photon detector initiative}

\subsection*{Total Cost (assuming no prior equipment)}
\$29,280 - \$29,780

\newpage

\bibliographystyle{unsrt}
\bibliography{hompaper}

\end{document}